\documentclass{revtex4-1}
\usepackage{graphicx}
\usepackage{bm}
\usepackage{amssymb,amsmath}
\usepackage{latexsym}
\newcommand{\G}{{G_F}}
\newcommand{\be}{\begin{equation}}
\newcommand{\ee}{\end{equation}}
\newcommand{\bea}{\begin{eqnarray}}
\newcommand{\eea}{\end{eqnarray}}
\newcommand{\gw}{{G^+}}
\newcommand{\D}{\mathcal{D}}
\newcommand{\K}{\mathcal{K}}

\begin{document}
\title{The hydrodynamic approximation of the semiclassical dissipation kernel 
in  stochastic gravity} 
\author{ Seema Satin } 
\affiliation{Indian Institute for Science Education and Research, Kolkata
 India} 
\email{seemasatin@iiserkol.ac.in }
\date{ }
\begin{abstract}
Semiclassical stochasic gravity is aimed at studying extended structure
 formation 
in the early universe. Rigorous developments in this area include the
 semiclassical
noise and dissipation kernels which are obtained in terms of quantum stress
energy tensor composed of scalar fields. The present article forms an 
important step in an effort to extend the theory in the decoherence limit
 and hydrodynamic approximation of the scalar fields. Such extensions
 will make it possible to analyse the extended structure formation around the 
decoherence era of the inflaton field in cosmology. On the other hand,
 modelling dissipation
 in fluids and effective fluids is a challenge  and long
standing mathematical physics related hurdles have posed difficulties for
 progress in this direction.
The present article marks the beginning of a new way to model dissipation in
an effective fluid using  the  widely accepted  correspondence between the
 semiclassical fields and effective fluid approximation. 
  A similar approach has been recently carried out for
noise (fluctuations) kernel  correspondence between the two, we now
touch upon  dissipation in the relativistic effective fluids.  
\end{abstract}
\maketitle
\section{Introduction}
The theory of semiclassical stochastic gravity is based on the semiclassical
Einstein-Langevin equation \cite{bei1}, where the fluctuations of quantum
 stress tensor 
act as noise to induce perturbations in the metric. The  two point noise
kernel composed of fluctuations of matter fields,  plays a
central role \cite{bei1,phillip,eftek}, in this theory. 
The  metric perturbations induced by the fluctuations of the matter fields 
can be used to  probe  extended structure in the early universe cosmology.
 However, the semiclassical Einstein 
Langevin equation in its full form needs to be solved in order to obtain these
results, which is a tedious task. The solutions for the semiclassical
Einstein-Langevin equation have been obtained for the Minkowski \cite{mink}
 and De-sitter spacetimes \cite{desit,desit1}, yet awaited are solutions for 
the FRW
 metric and other cosmologically relevant spacetime configurations. Not just 
the noise kernel expressions
but also dissipation kernel and  more  terms need to be evaluated explicitly
 for a full solution of the semiclassical E-L (Einstein-Langevin) equation 
which reads \cite{bei1},
\be \label{eq:EL}
G_{ab}[g+h] + \Lambda(g_{ab} + h_{ab} ) - 2(\alpha A_{ab}+ \beta B_{ab})[g+h]
= 8 \pi G( \langle \hat{T}^R_{ab}[g+h] \rangle  + \tau_{ab}[g])
\ee
where $\tau_{ab}$ represents the noise term on the background spacetime
metric $g_{ab} $. 
The perturbations of the  regularized stress tensor, $\hat{T}^R_{ab}[g+h] $ in
 the above, includes dissipation kernel, which is the focus  in the present
 article.  

One of our long term goals in the research program is to obtain and
solve the Einstein-Langevin equation around the decoherence era in 
early universe. 
Structure formation in this epoch, around the decohernce era, is open for
 investigation and holds promises for exciting new developments in theory and
for  observations.  We intend to work on foundations and a rigorous theoretical
 base for this. Important significant mathematical results and developments 
in overlapping
areas like relativistic and non-relativistic fluids are expected to follow
in due course.    
 
The dissipation kernel  along with the noise kernel in semiclassical stochastic gravity program 
 has been obtained through an influence functional and using closed time path
(CTP)  formalism \cite{martin,bei3}.
 The quantum fields and quantum open systems approach cannot be directly
extended to classical systems to account for dissipation, though recently
there have been prelimiary  efforts which are limited to a very simple case
\cite{chad}. But
 the correspondence in terms of the stress energy tensor  and scalar 
fields in the classical limit follows easily through the first principles.
We use this later approach to find a classical  hydro limit approximation
 for the dissipation kernel in this article. This is a foundational level 
building block in the research program.

In the following section we give a brief review of the semiclassical
dissipation kernel.  
\section{Review of the semiclassical dissipation kernel} 
Elaborate formalism to obtain  dissipation kernel in the semiclassical
Einstein-Langevin equation from the second order
expansion of the influence action and closed time path formalism are
given in \cite{bei1,martin,bei2}. We give below  the final forms of expressions
and  will utilze them  show correspondence with the
fluid approximation of the fields.   
Regularization procedure for quantum fields is elaborately taken care of in
 the formulation
of the semiclassical theory. 
The quantum stress tensor for the  (scalar) inflaton field in cosmology reads,
\be
\hat{T}^{ab} (x) = \frac{1}{2} \{ \nabla^a \hat{\phi}[g],\nabla^b \hat{\phi}
[g] \} + \mathcal{D}^{ab} [g] \hat{\phi}^2 [g] 
\ee
where $\mathcal{D}^{ab}[g] $,
\be
\mathcal{D}^{ab} = (\xi - 1/4) g^{ab} \Box + \xi (R^{ab} - \nabla^a \nabla^b)
-\frac{1}{2} m^2 g^{ab}
\ee

In equation (\ref{eq:EL}) the term $T_{ab}[g+h]$ can be evaluated as
\bea
\hat{T}^{ab}(x)[g+h,\hat{\phi}]= \hat{T}^{ab}[g,\hat{\phi}] + \langle 
\hat{T}^{(1)ab}[g;h](x) \rangle [g] - 2 \int d^4 y  \sqrt{- g(y)} H^{abcd}
[g](x,y) h_{cd}(y)
\eea
where $H^{abcd}[g](x,y)$ is the dissipation kernel is obtained from the second
 order
 expansion in the influence functional method \cite{bei1,bei3}. 
 This kernel can written as  $ H^{abcd}(x,y) = H^{abcd}_A(x,y) + 
H^{abcd}_S(x,y) $, where the
 antisymmetric part  $ H^{abcd}_A (x,y) = - H^{abcd}_A(y,x)$ and the symmetric
part $ H^{abcd}_S (x,y) = H^{abcd}_S (y,x)$, which are given by,
\bea \label{eq:ha}
& & H^{abcd}_A(x,y) =  \frac{i}{4} \langle \frac{1}{2} [\hat{T}^{ab}(x),
\hat{T}^{cd}(y)]\rangle [g]  \\
& & H^{abcd}_S (x,y) = \frac{1}{4} \mbox{Im} \langle \mathbf{T^*}(\hat{T}^{ab}
(x) \hat{T}^{cd}(y)) \rangle [g] \mbox{ where $\mathbf{T^*}$ is the time
ordering operator } \label{eq:hb} 
\eea
We will utilize this form of the expressions to obtain the fluid 
correspondence.

 The explicit expressions for the dissipation kernel \cite{bei1} using the
quantum stress tensor for the scalar fields, and 
  (\ref{eq:ha}) and (\ref{eq:hb}) have been elaborately obtained as: 
\bea \label{eq:haf}
H_A^{abcd}(x,y) & = & \nabla^a_x \nabla^c_y  \gw \nabla^b_x \nabla^d_y \gw +
 \nabla^a_x \nabla^d_y \gw \nabla^b_x \nabla^c_y \gw + \nonumber \\
& &  2 \D^{ab}_x ( \nabla_y^c \gw \nabla^d_y \gw + 
2 \D^{cd}_y ( \nabla^a_x \gw \nabla^b_x \gw ) + 2 \D_x^{ab} \D_y^{cd} (\gw^2)
\eea
where $\gw \equiv \gw(x,y)$ are the Wightman functions $\gw(x,y) = \langle
 0 |\hat{\phi}[g;x) \hat{\phi}[g;y)|0 \rangle $. And,   
\bea \label{eq:hbf}
H^{abcd}_s(x,y) & = & - \mbox{Im} [  \nabla_x^a \nabla_y^c \G  \nabla_x^b 
\nabla_y^d  \G	 + \nabla^a_x \nabla^d_y \G \nabla^b_x \nabla^c_y \G
-g^{ab}(x) \nabla^e_x \nabla^c_y \G \nabla_e^x \nabla_y^d \G  \nonumber \\
& & 
- g^{cd}(y) \nabla_x^a \nabla_y^e \G \nabla_x^b \nabla_e^y \G
+ \frac{1}{2} g^{ab}(x) g^{cd}(y) \nabla^e_x \nabla^f_y \G 
\nabla^x_e \nabla^y_f \G +  \nonumber \\
& & \K_x^{ab} ( 2 \nabla^c_y \G \nabla^d_y \G - 
g^{cd}(y) \nabla^e_y \G \nabla^y_e \G + 2 \K^{ab}_x \K^{cd}_y
(\G^2)]
\eea
where  $\G \equiv \G(x,y) $ are the Feynman functions $ i \G(x,y) = \langle
0 |\mathbf{T} \hat{\phi} [g;x) \hat{\phi}[g;y)|0 \rangle $, $\mathbf{T}$ being
 the time ordering operator, and where 
\be
 \K^{ab}_x  =  \xi ( g^{ab} (x) \Box_x - \nabla^a_x \nabla^b_x
+ G^{ab}(x)) - \frac{1}{2} m^2 g^{ab} (x)  
\ee
In the decoherence limit and classical approximation $\hat{\phi}[g;x) 
\rightarrow \phi[g;x) $ and expectation $\langle \hat{\phi}(x) \rangle $
can be approximated with $\int \phi (x) P(\phi) \mathcal{D} \phi $ where
$P(\phi)$ denotes a probability distribution and $\mathcal{D} \phi$ denotes
a functional integral. Then the two point function given by
$G^+(x,x')$  and $G_F(x,x') $ can be approximated by $\int \phi(x) \phi(x')
P(\phi) \mathcal{D} \phi $. This has also been discussed in 
details in \cite{satin1,satin2}.   
\section{Dissipation kernel in the hydrodynamic approximation} 
In this section we work out the main results for the hydrodynamic
approximation of the  the semiclassical two point dissipation kernel composed
 of scalar fields.  
 It is known that  the quantum field stress tensor  $\hat{T}^{ab}(x)$ has 
 correspondence with the fluid stress tensor $T^{ab}(x)$ in the hydrodynamic
limit \cite{madsen}, where 
\bea
\hat{T}^{ab(field)}(x) = \frac{1}{2} \{\nabla^a \hat{\phi}[g] , \nabla^b \hat{\phi}[g]
\} + (\xi - 1/4) g^{ab}\Box \hat{\phi}^2 + \xi(R^{ab} - \nabla^a \nabla^b)
\hat{\phi}^2 - \frac{1}{2} m^2 g^{ab} \hat{\phi}^2
\eea
corresponds to 
\bea \label{eq:stressf}
T^{ab(fluid)} = u^a u^b (\epsilon+ p) + g^{ab} p + q^a u^b + u^a q^b + 
\pi^{ab} 
\eea
such that $u_a$ represents the four velocity, $\epsilon$ the energy
density, $p$ as pressure of the fluid, $q_a $ heat flux, $\pi_{ab} $ 
anisoptropic stresses.

Where taking $V(\phi) = \frac{1}{2} m^2 \phi^2 $, one can relate the
hydrodynamic variables with the scalar field as,

\bea
u_a & = &  [\partial_c (\phi) \partial^c (\phi) ]^{-1/2} \partial_a
\phi  \label{eq:vel}\\
\epsilon & =& [ \frac{1}{2} \partial_c \phi \partial^c \phi + V(\phi) + 
\xi \{\Box(\phi^2) - (\partial^c \phi \partial_c \phi )^{-1}
\partial^a \phi \partial^b \phi \nabla_a \nabla_b ( \phi^2) \}] \label{eq:ep}\\
q_a &= & \xi ( \partial^b \phi \partial_b \phi ) ^{-3/2}
\partial^c \phi \partial^d \phi [ \nabla_c \nabla_d (\phi^2) \partial_a \phi
- \nabla_a \nabla_c (\phi^2) \partial_d \phi ] \label{eq:q}\\
p & = &  [ \frac{1}{2} \partial_c  \phi \partial^c
 \phi - V(\phi) -  \xi \{ \frac{2}{3} \Box (\phi^2) + \frac{1}{3} (\partial_c
 \phi \partial^c \phi )^{-1} \nabla_a \nabla_b (\phi^2) \partial^a \phi
\partial^b \phi \} ] \label{eq:p} \\
\pi_{ab} & = & \xi  ( \partial^c \phi \partial_c \phi)^{ -1} [
\frac{1}{3} ( \partial_a \phi \partial_b \phi - g_{ab} \partial^c \phi
 \partial_c \phi ) \{ \Box ( \phi^2) - (\partial^c \phi \partial_c \phi )^{-1}
\nabla_c \nabla_d (\phi^2) \partial^c \phi \partial^d \phi \} \nonumber \\
& &   + \partial^p \phi \{ \nabla_a \nabla_b ( \phi^2) \partial_p \phi -
\nabla_a \nabla_p (\phi^2) \partial_b \phi - \nabla_p \nabla_b
( \phi^2) \partial_a \phi + ( \partial_c \phi \partial^c \phi ) ^{-1}
\partial^d \phi \nabla_d \nabla_p (\phi^2) \partial_a \phi \partial_b \phi \}]
\nabla_a \nabla_p (\phi^2) \partial_b \phi  \nonumber \\
& &- \nabla_p \nabla_b ( \phi^2) \partial_a \phi + ( \partial_c \phi
\partial^c \phi ) ^{-1} \partial^d \phi \nabla_d \nabla_p (\phi^2)
\partial_a \phi \partial_b \phi \}]
\label{eq:pi}
\eea
 
 The hydro approximation is taken in the classical limit, hence
one can  take  a step inbetween, $\hat{\phi}[g] \rightarrow \phi[g]$,
 a classical scalar field
 and use the corresponding stress tensor $\hat{T}^{ab(field)} \rightarrow
T^{ab(field)}$ for the classical stress tensor, which then is shown to have
a correspondence with the fluid stress tensor $T^{ab(fluid)} $  in the hydro
approximation. 

The dissipation kernel in terms of the stress tensor takes the form, 
\bea
& & H^{abcd(field)}_A(x,y) =  \frac{i}{4} \langle \frac{1}{2} [\hat{T}^{ab}(x),
\hat{T}^{cd}(y)]\rangle [g]  \rightarrow  
\frac{1}{4} \langle \frac{1}{2} \{T^{ab}(x), T^{cd}(y)\}\rangle [g] = 
H^{abcd(fluid)}(x,y) \label{eq:anti} \\
& & H^{abcd(field)}_S (x,y) = \frac{1}{4} \mbox{Im} \langle \mathbf{T^*}(
\hat{T}^{ab}(x) \hat{T}^{cd}(y)) \rangle [g] \rightarrow  
 \frac{1}{4} \mbox{Im} \langle ({T}^{ab}(x)T^{cd}(y) \rangle = 
H^{abcd(fluid)}_S(x,y).
\eea
 where $\{,\}$ in equation  (\ref{eq:anti}) denotes Poisson
bracket. 
\subsection{Antisymmetric part of dissipation kernel in the hydrodynamic
approximation}
The antisymmetric  part in the semiclassical case which is the 
commutator of quantum stress tensors has  an elaborate 
form  given as
\bea
H^{abcd(field)}_A(x,y) &= & \nabla^a_x \nabla^c_y  \gw \nabla^b_x \nabla^d_y
 \gw + \nabla^a_x \nabla^d_y \gw \nabla^b_x \nabla^c_y \gw + 
 2 \D^{ab}_x ( \nabla_y^c \gw \nabla^d_y \gw + 
2 \D^{cd}_y ( \nabla^a_x \gw \nabla^b_x \gw ) + 2 \D_x^{ab} \D_y^{cd} (\gw^2)
\eea  

The fluid approximation of the  
antisymmetric dissipation is the Poisson bracket  for classical 
 stress tensors $\{T^{ab(fluid)}(x),T^{cd(fluid)}(y) \}$.
Evaluating this Possion bracket  explicity (and also commutators of the
 quantum stress tensor) is an open
research problem in mathematical physics, for which very few results have been
known till now \cite{forger}. With our endeavor to define dissipation in 
the fluid approximation of fields, this gains more importance and motivation.
In this  article, we present the important first explicit evaluation.  

The Poisson bracket $ \{ , \} $ has to be evaluated with the symplectic
structure for the stress tensor, $T^{ab}(q,p)$, where $q,p$ represent
 generalized coordinate
 and momenta. For the fluid stress tensor given in terms of fluid
variables such a symplectic structure has not been constructed yet, and it
is a non-trivial mathematical physics problem. We find an alternative and
 feasible way to evalute this. Limiting ourselves to flat spacetime,
 for a perfect fluid and exploiting the scalar field-fluid
correspondence it is possible to get analytical closed form expressions. 
 We leave the more involved case for an imperfect fluid  for  
 for future work.     

\textbf{Evaluation for Minkowski spacetime}

Taking the form of stress tensor for the classical scalar field as 
\be
 T^{ab} =  \phi_{;a} \phi_{;b} + \frac{1}{2} \eta_{ab}( \phi^{;c} \phi_{;c}
+ m^2 \phi^2) 
\ee
the evaluation for the Poission bracker $\{T_{ab}(x), T_{cd}(y)\}$ needs a
 symplectic structure to be defined. We proceed as follows.

For flat spacetime ( Minkowski with $ \eta_{ab} \equiv (-1,1,1,1)$)  
using a "."  to denote time derivate we
 define the conjugate variables as $\phi$ and $ \pi = \dot{\phi}$, such that
 the canonical Poission brackets
are 
\bea
\{\phi(x), \phi(y)\} = 0 ,  \{\pi(x), \pi(y) \} = 0 & & 
\nonumber\\
\{\phi(x),\pi(y) \} = \delta(x-y)
\eea   
 With this we  obtain for 
\be \label{eq:po}
 \{T_{ab}(x),T_{cd}(y) \} 
\ee
the following results in a lengthy but straightforward 
evaluation,
\bea \label{eq:pois}
& &  \{T_{00}(x) T_{00}(y) \} = (T_{0\alpha}(x) + T_{0\alpha}(y) )
 \partial^\alpha \delta(x-y)
\nonumber \\
& & \{T_{00}(x), T_{0\gamma}(y) = T_{00}(y) \partial_\gamma \delta(x-y) +
 T_{\gamma \delta}(x) \partial^{\delta} \delta(x-y) \nonumber \\
& & \{T_{0 \alpha}(x), T_{0 \beta}(y) \} = T_{0 \beta}(x) \partial_\alpha 
\delta(x- y) + T_{0 \alpha}(y) \partial_\beta \delta(x-y) \nonumber \\
& & \{T_{\alpha \beta}(x), T_{\gamma \delta}(y) \}  = 0 
\eea
We can easily make a correspondence between the above scalar field
results and the fluid approximation, by replacing rhs of the above set
of equations  with $T^{ab(fluid)}$.
Thus we have devised a new way for obtaining the Poission bracket of
fluid stess tensor using the scalar field correspondence and a
related  symplectic structure. The correspondence for the antisymmetric
part of the dissipation kernel then can be presented in separate
 components as 
\bea
& & H_A^{0000}(x,y) = \nabla^0_x \nabla^c_y  \gw \nabla^0_x \nabla^0_y \gw +
 \nabla^0_x \nabla^0_y \gw \nabla^0_x \nabla^0_y \gw + 
 2 \D^{00}_x ( \nabla_y^0 \gw \nabla^0_y \gw + 
2 \D^{00}_y ( \nabla^0_x \gw \nabla^0_x \gw ) + 2 \D_x^{00} \D_y^{00} (\gw^2)
\rightarrow \nonumber \\
& &   (u^0(x) u^\gamma(x)( \epsilon(x) + p(x)) +
  u^0(y) u^\gamma(y) (\epsilon(y) + p(y)) ) \partial^\gamma \delta(x-y) 
\eea
\bea
& & H_A^{000\gamma}(x,y)= \nabla^0_x \nabla^0_y  \gw \nabla^0_x
 \nabla^\gamma_y \gw + \nabla^0_x \nabla^\gamma_y \gw \nabla^0_x \nabla^0_y
 \gw + 2 \D^{00}_x ( \nabla_y^0 \gw \nabla^\gamma_y \gw + 
2 \D^{0 \gamma}_y ( \nabla^0_x \gw \nabla^0_x \gw ) + 2 \D_x^{00} 
\D_y^{0 \gamma} (\gw^2)
\rightarrow \nonumber \\
& & (u^0(y))^2 (\epsilon(y) + p(y)) - p(y)) \partial^\gamma \delta(x-y) +
 (u^\gamma(x) u^\delta(x)(\epsilon(x) + p(x) )+ \eta^{\gamma \delta} p(x))
 \partial_\delta \delta(x-y) 
\eea
\bea
& & H_A^{0 \alpha 0 \beta}(x,y)= \nabla^0_x \nabla^0_y  \gw \nabla^\alpha_x 
\nabla^\beta_y \gw + \nabla^0_x \nabla^\beta_y \gw \nabla^\alpha_x 
\nabla^0_y \gw + 
 2 \D^{0 \alpha}_x ( \nabla_y^0 \gw \nabla^\beta_y \gw + 
2 \D^{0 \beta}_y ( \nabla^0_x \gw \nabla^\alpha_x \gw ) + 2 \D_x^{0 \alpha}
 \D_y^{0 \beta} (\gw^2)
\rightarrow \nonumber \\
 & & u^0(x) u^\beta(x)(\epsilon(x) + p(x)) \partial^\alpha \delta(x-y) +
 u^0(y) u^\alpha(y)(\epsilon(y) + p(y)) \partial^\beta \delta(x-y)  
\eea  
 and
\bea
& & H_A^{\alpha \beta \gamma \delta}(x,y)= \nabla^\alpha_x \nabla^\gamma_y  \gw 
\nabla^\beta_x \nabla^\delta_y \gw +
 \nabla^\alpha_x \nabla^\delta_y \gw \nabla^\beta_x \nabla^\gamma_y \gw + 
 2 \D^{\alpha \beta}_x ( \nabla_y^\gamma \gw \nabla^\delta_y \gw +  \nonumber\\
& &
2 \D^{\gamma \delta}_y ( \nabla^\alpha_x \gw \nabla^\beta_x \gw ) + 2
 \D_x^{\alpha \beta} \D_y^{\gamma \delta} (\gw^2)
\rightarrow 0   
\eea  

\textbf{For Curved Spacetime :}

A general prescription for Poisson bracket in curved spacetime is known to 
take the form \cite{wald}, which for the stress tensors can be written
in the form
\be
\{T^{ab},T^{cd} \} = 
\Omega^{ef} \nabla_e T^{ab} \nabla_f T^{cd}
\ee 
 where $\Omega^{ab}(x,y)$ needs to be defined at $(x,y)$ and is defined as the
symplectic structure on the spacetime manifold.
\be
\Omega_{ab} = 2 \sum_\mu (\nabla p_{\mu})_{[a}(\nabla q_{\mu})_{b]}
\ee
where $p,q$ are local conjugate coordinates and momenta on the manifold.  
 
This is just a proposed ansatz and elaborate evaluation needs rigorous 
mathematical developments which is our future plan.
\subsection{Symmetric part of the dissipation kernel in hydrodynamic 
approximation}
For the symmetric part of the dissipation kernel,  the fluid approximation
is again given by simply taking $T^{ab(field)} \rightarrow T^{ab(fluid)} $,
 for which the expression reads, 
\bea \label{eq:sfluid}  
H^{abcd(fluid)}_S(x,y) = 
& & \frac{1}{2}\{ \{ \langle u_a(x) u_c(y) \rangle \langle u_b(x) u_d(y)
 \rangle + \langle u_a(x) u_d(y) \rangle \langle u_b(x) u_c(y) \rangle \}
\{ \langle \epsilon(x) \epsilon(y) \rangle + \nonumber \\
& &  \langle \epsilon(x) p(y) \rangle
+ \langle p(x) \epsilon(y) \rangle + \langle p(x) p(y) \rangle \}
+ \langle q_a (x) q_c(y)  \rangle \langle u_b(x) u_d(y) \rangle + 
\langle q_a(x) q_d(y) \rangle \langle u_b(x) u_c(y) \rangle 
+ \nonumber \\
& & \langle u_a(x) u_d(y) \rangle \langle q_b(x) q_c(y) \rangle + 
\langle u_a(x) u_c(y) \rangle \langle q_b(x) q_d(y) \rangle + \langle \pi_{ab}
\pi_{cd}(y) \rangle \}
+ g_{ab}(x) \{\langle u_c(y)\rangle \langle u_d(y) \rangle \{ \langle p(x)
 \epsilon(y) \rangle \nonumber\\
& &  + \langle p(x) p(y) \rangle \} \}
+ g_{cd}(y) \{ \langle u_a(x) \rangle \langle u_b(x) \rangle \{ \langle 
\epsilon(x) p(y) \rangle + \langle p(x) p(y) \rangle  \} \}
+ g_{ab}(x) g_{cd}(y) \{ \langle p(x) p(y) \rangle \}
\eea
We now show the  term by term correspondence, between the fluid and 
field dissipation kernels (symmetric part)  comparing expressions
 (\ref{eq:hbf}) and (\ref{eq:sfluid}) for which we  separate terms as
 coefficients of
 $g_{ab}$ metric factors in the expressions. With the coefficient of
 $g^{ab} $,
\bea
& &  -\G^{;ce} \G^{;d}_{;e} + 2 \xi \Box_x \G^{;c} \G^{;d} -
 m^2 \G^{;c} \G^{;d} - (\G^{;e} \G_{;e} )^{;cd} - \nonumber \\
& & \G^{;e} \G_{;e} G^{cd} - \xi ( \G^{;e} \G_{;e})^{;cd} - \G^{;e} \G_{;e}
 G^{cd} - \xi^2 {(\Box_x \G^2)}^{;cd} - \nonumber \\
& & \xi^2 G^{cd} \Box_x \G^2 - \frac{1}{2} m \xi (\G^2)^{;cd} -
\frac{1}{2} m^2 \xi G^{cd} \G^2   \rightarrow 
\langle u_c(y)\rangle \langle u_d(y) \rangle \{ \langle p(x) \epsilon(y)
 \rangle + \langle p(x) p(y) \rangle \} 
\eea

Coefficient of $g^{cd}$,

\bea
 & & - \G^{;ea} \G_{;e}^b - \xi (G^{;e} \G_{;e} )^{;ab} - \xi \G^{ab} \G^{;e} 
\G_{;e} + 2 \xi \Box_x \G^{;a} \G^{;b} \nonumber \\
& & 2 \xi \Box_x \G^{;a} \G^{;b} - m^2 \G^{;a} \G^{;b} - \xi^2 \Box_x \Box_y 
\G_F^2 + \xi^2 \G^{ab} \Box_y \G^2 \nonumber \\
& & + \xi^2 G^{ab} \Box_y G_F^2 - \xi m^2 (\G^2)^{;ab} + \xi m^2 G^{ab}\G^2
\rightarrow 
\langle u_a(x) \rangle \langle u_b(x) \rangle \{ \langle \epsilon(x) p(y)
 \rangle + \langle p(x) p(y) \rangle  \}
\eea
coefficient of $ g^{ab} g^{cd} $ 
\bea
& & \frac{1}{2} \G^{;fe} \G_{;fe} - \xi^2 \Box_x \G^{;e} \G_{;e} - 
- \xi \Box_y \G^{;e} \G_{;e}  \nonumber \\
& & - \xi \Box_y (\G_{;e} \G^{;e}) + 2 \xi^2 \Box_x \Box_y
\G^2 - \xi m^2 \Box_x \G^2 - \frac{1}{2} m^2 \xi \Box_y \G^2 \nonumber \\
& & + \frac{1}{2} m^2 \G^2 \rightarrow 
\langle p(x) p(y) \rangle \rightarrow \frac{1}{2} \G^{;fe} \G_{;fe} 
- \frac{1}{2} m^2 \G^{;e} \G_{;e} -\frac{1}{2} m^2 \G^2
\eea
And,
\bea 
& & \G^{;ca} \G^{;db} + \G^{;da} \G^{cb} - ( \G^{;c} \G^{;d})^{;ba} - 
2 \xi G^{ab} \G^{;c} \G^{;d} - \nonumber \\
& & 2 \xi (G^{;d} \G^{;b})^{;cd} + 2 G^{cd} \G^{;a} \G^{;b} -
 2 ( \G^{;d} \G^{;b} )^{;cd} - 2 G^{cd} \G^{;a}
\G^{;b} \nonumber \\
& &  - \xi^2 (\G^2)^{;dcba} + \xi^2 G^{cd} (\G^2)^{;ba} + \xi^2 G^{ab} 
(\G)^{;cd} + \xi^2 G^{ab} G^{cd} \G^2. \rightarrow \\
& & \frac{1}{2}\{ \{ \langle u_a(x) u_c(y) \rangle \langle u_b(x) u_d(y)
 \rangle + \langle u_a(x) u_d(y) \rangle \langle u_b(x) u_c(y) \rangle \}
\{ \langle \epsilon(x) \epsilon(y) \rangle + \nonumber \\
& &  \langle \epsilon(x) p(y) \rangle
+ \langle p(x) \epsilon(y) \rangle + \langle p(x) p(y) \rangle \}
+ \langle q_a (x) q_c(y)  \rangle \langle u_b(x) u_d(y) \rangle + 
\langle q_a(x) q_d(y) \rangle \langle u_b(x) u_c(y) \rangle 
+ \nonumber \\
& & \langle u_a(x) u_d(y) \rangle \langle q_b(x) q_c(y) \rangle + 
\langle u_a(x) u_c(y) \rangle \langle q_b(x) q_d(y) \rangle + 
\langle \pi_{ab}(x) \pi_{cd}(y) \rangle \}
\eea
The symmetric part of the dissipation kernel is easy to evaluate for a general
imperfect fluid as we see above. 
For $\xi = 0 $, the  approximation reduces to the perfect fluid,
where heat flux $q_a $ and $\pi_{ab} $ anisotropic stresses are zero. Thus 
 
\bea
u_a & = &  [\partial_c (\phi) \partial^c (\phi) ]^{-1/2} \partial_a
\phi  \label{eq:vel1}\\
\epsilon & =&  \frac{1}{2} \partial_c \phi \partial^c
 \phi + V(\phi)  \label{eq:ep1}\\
p & = &   \frac{1}{2} \partial_c  \phi \partial^c \phi - V(\phi) \label{eq:p1}
\eea
while the scalar field stress tensors reduce to,
\bea
\hat{T}^{ab(field)}(x) = \frac{1}{2} \{\nabla^a \hat{\phi}[g] , \nabla^b 
\hat{\phi}[g] \}  - 1/4 g^{ab}\Box \hat{\phi}^2 
\eea
and  for the fluid as, 
\bea
T^{ab(fluid)} = u^a u^b (\epsilon+ p) + g^{ab} p  
\eea

Than for the symmetric part of the dissipation kernel 
for perfect fluid approximation, we have the correspondence : 

given by coefficient of $g^{ab}$,
\bea
& &  - \G^{;ce} \G^{;d}_{;e} - m^2 \G^{;c}
\G^{;d} - (\G^{;e} \G_{;e})^{;dc} - \G^{;e} \G_{;e} \G^{;cd} \rightarrow 
\nonumber \\
& & \langle u^c(y) \rangle \langle u^d(y) \rangle \{ \langle p(x) \epsilon(y)
 \rangle + \langle p(x) p(y) \rangle \}  \nonumber \\
\eea

coefficient of $g^{cd} $.
\bea
& & - \G^{;ea} \G_{;e}^{;b} - \frac{1}{2} m^2 \G^{;a} \G^{;b} \rightarrow
\nonumber \\
& & \langle u^a(x) \rangle \langle u^b(x) \rangle \{ \langle \epsilon(x) p(y)
 \rangle + \langle p(x) p(y) \rangle \}  \\
\eea

coefficient of $g^{ab} g^{cd} $,

\bea
\frac{1}{2} \G^{;fe} \G_{;fe} - \frac{1}{2} m^2 \G^{;e} \G_{;e} -
\frac{1}{2} m^2 \G^2 \rightarrow \langle p(x) p(y) \rangle  
\eea
and 
\bea
& & \G^{;ca} \G^{;db} + \G^{;da} \G^{;cb} - (\G^{;e} \G^{;d})^{;ba}
+ 2 G^{cd} \G^{;a} \G^{;b} -  \nonumber \\
& & 2 ( \G^{;d} \G^{;b} )^{;dc} - 2 G^{cd} \G^{;a} \G^{;b} \rightarrow \\
& & \{\langle u^a(x) u^{(c}(y) \rangle \langle u^b(x) u^{d)}(y) \} 
\{ \langle \epsilon(x) \epsilon(y) \rangle + \langle \epsilon(x) p(x) 
\rangle + \langle p(x) \epsilon(x) \rangle 
 + \langle p(x) p(y) \rangle \} 
\eea
For this case,  when  $\xi =0 $,  one can also work out the
reverse relations from the above, giving one to one correspondence
\bea
& &  \langle p(x) p(y) \rangle \rightarrow \frac{1}{2} \G^{;fe} \G_{;fe} -
 m^2 \G^{;e} \G_{;e} - \frac{1}{2} m^2 \G^2   \\
& & \langle \epsilon(x) p(y) \rangle \rightarrow - \langle \frac{ [\partial_e 
\phi(x) \partial^e \phi(x)]^{1/2} \partial_f \phi(x) \partial^f 
\phi(x) ]^{1/2} }{ \partial_a \phi(x) \partial_b \phi(x) } \rangle
 [ - \G^{;ea} \G{;e}^{;b} - \frac{1}{2} m^2  \G^{;a} \G^{;b} ] \nonumber \\
& & - \frac{1}{2}  \G^{;fe} \G_{;fe} + \frac{1}{2} m^2 \G^{;e} \G_{;e} 
+ \frac{1}{2} m^2 \G^2  \\
& & \langle p(x) \epsilon(y) \rangle \rightarrow - \langle \frac{[\partial_e
 \phi(y) \partial^e \phi(y) ] }{\partial_c \phi(y) \partial_d \phi(y) }
\{ \G^{;ce}\G^{;d}_{;e} - m^2 \G^{;c} \G^{;d} -  (\G^{;e} \G_{;e})^{;dc} 
\nonumber \\
& & -\G^{;e} \G_{;e} \G^{;cd} \} -\frac{1}{2} \G^{;fe} \G_{;fe} -
 \frac{1}{2} m^2 \G^{;e} \G_{;e} - \frac{1}{2} m^2  \G^2  \\
& & \langle \epsilon(x) \epsilon(y) \rangle \rightarrow  \langle \frac{
 [\partial_e \phi(x) \partial^e \phi(x) ][\partial_f \phi(y) \partial^f
 \phi(y) ]}{ \partial_a \partial_{(c} (\phi(x) \phi(y)) \partial_b
 \partial_{d)} 
\phi(x) \phi(y)} \rangle [ \G^{;ca} \G^{;db} + \G^{da} \G^{cb} - ( \G^{;c}
\G^{;d})^{;ba} + \nonumber \\
& & 2 G^{cd} \G^{;a} \G^{;b} - 2 ( \G^{;a} \G^{;b} )^{;dc} 
- 2 G^{cd} \G^{;a} \G^{;b} ] - \langle \frac{[\partial_e \phi(x) \partial^e 
\phi(x) ]}{\partial_a \phi(x) \partial_b \phi(x) } \rangle \nonumber \\
 & & [- \G^{;ea} \G_{;e}^{;b} - \frac{1}{2} m^2 \G^{;a} \G^{;b} ] +
\frac{1}{2} \G^{;fe} \G_{;fe} - \frac{1}{2} m^2 \G^{;e} \G_{;e} \nonumber \\
& &  - \frac{1}{2} m^2 \G^2
\eea
 In this section we have presented a complete picture of the symmetric
part of the dissipation kernel, for both the imperfect and perfect 
relativistic fluids.
\section{The fluctuation-dissipation relation in the hydrodynamic limit}
 By  combining the expressions for the
fluctuations in terms of the  noise kernel worked out in
\cite{satin1,satin2,satin3} and the dissipation kernel, obtained in this 
article, one can write a fluctuation-
dissipation relation in the hydro limit. It is to be noted  that
the mathematical structure for noise and dissipation in the classical 
approximation using our approach follows from taking limit of the 
quantum fields and fluctuations. However the physical origin of quantum
fluctuations and dissipation  is very different from the classical origin. The
question  to pose is, what is the underlying physical nature of these classical
approximation fluctuations and dissipation. A general understanding in the
theory of fluctuations has been , these should be thermal if nor quantum. The
magnitude of thermal fluctuations in hydro approximation can be very small
specially in the equilibrium states and hence can be considered to be
 insignificant. However, there is 
another physical origin to noise and dissipation, which is not very commonly
acknowledged. Non-thermal, purely mechanical effects including linear
as well as non-linearity in the fluids or any complex mechanical system 
can give rise to dynamical fluctuations and  dissipate
energy by adiabatically perturbing the system. Similar well
known phenomena is turbulence in ordinary and relativistic fluids. We wish to
 highlight such a 
physical origin  for purely mechanical fluctuations and dissipation in our
 research program. 
This has good scope for new developments with openings for theoretical 
research and applications.

From the semiclassical theory of stochastic gravity, it is
known that the antisymmetric part of the dissipation kernel is
linked to the noise kernel via a fluctuation dissipation relation \cite{martin,
sukanya}. 
For the classical limit in the hydro approximation, the noise kernel worked
 out in \cite{satin1,satin2} can be related to the
antisymmetric part of the dissipation kernel discussed in this article through
a fluctuation dissipation relation. Such a relation for the semiclassical 
theory with quantum stress tensors is well established \cite{sukanya}, a 
similar form
can be written down for the hydrodynamic approximation as
\be
 N^{abcd} (x,x') = \int K(x',x'') H^{abcd}_A (x'',y) dx''
\ee
where the FD kernel $ K(x',x'') $ is defined on the background geometry and 
relates the noise and dissipation kernels.
Thus if we know the Poisson bracket for the fluid stress tensor given by 
(\ref{eq:po}), we can present the above relation in the form
\be
 N^{abcd} (x,x') = \int K(x',x'') \{T^{ab}(x''),T^{cd}(y) \} dx''
\ee
The above relation is a general one and valid for flat as well
as curved spacetimes.
\section{Conclusion and further directions}
In this article we have obtained new expressions for dissipation in effective
fluid, using the scalar field-fluid
correspondence. This has been worked out using the decoherence/
classical limit for the scalar fields and using the stress energy tensor as
 the main entity to describe the fields and fluid on a spacetime manifold.

 The antisymmetric part of the dissipation kernel is shown to be represented
 by Poisson bracket of the
stress tensor in the classical limit, which is an important basic step in
our understanding of how dissipation can be modelled in a classical hydrodynamic
 theory for the relativistic  case. The explicit expressions for a general
case is a tedious task to evaluate and poses mathematical challenges. We have
 presented our first  case here which is
simpler to evaluate, that for 
the perfect fluid in flat spacetime. The next  that one should consider
to evaluate for, is the imperfect fluid approximation and the explicit 
Poission bracket details. The more involved and
open direction for mathematical physics developments is the explicit
evaluation of a Poission bracket  of the stress energy tensors for different
 cases of fields to fluid transition. We have also presented the general form
 for the Poisson bracket of the stress tensors in curved spacetime. Further 
work is in progress. 

 The antisymmetric part of the dissipation kernel can be
easily related to  noise  and this is presented in the previous section as
the fluctuation-dissipation relation, which follows easily from the
well  known semiclassical relation. The symmetric part of the dissipation
easily follows from the stress energy hydro aproximation. Both the symmetric
and antisymmetic parts  have been component wise related
in terms of scalar fields an hydrodynamic variables. These details are
of significance to explore the  underlying nature of  effective
fluid and hydro variable  correlations in a statistical physics
study.The equilibrium as well as non-equilibrium
phenomena then can be probed and analysed  starting with these results
 for any relativistic fluid.

 It is also interesting to see that we have not needed any specific
physical arguments to account for dissipation ( or noise) in our  approach, in
the sense that we do not worry about thermal  effects and 
 concerns to relate with temperature. Thus, these results are also valid for
non-thermal and cold matter perfect fluids, which we also infer from our
results. It is then the purely dynamical and mechanical effects that
can give rise to fluctuations and non-thermal dissipation (in classical systems)
 where energy
 may be dissipated in a different way than thermal losses. An example
is that of inducing adiabatic perturbations in the system where
the energy is dissipated in driving or inducing dynamical (mechanical)
 perturbations instead of heat loss. Such
 new physical ideas are interesting to consider in fluid dynamics as well as 
in relativistic astrophysics  for formulating  sub-hydro  multi
 scale mesoscopic theories for the systems. 
 
However, this article mainly is for the purpose of beginning new efforts to
 mathematically model 
dissipation in relativistic fluids and trying to understand the decoherence
limit correspondences between the fields and fluids at sub-hydro mesoscopic
 scales.

\textbf{Data Availability Statement:} No data has been generated in this work.

\section*{Acknowledgements}
This work has been funded through grant number DST WoSA/PM3/2021, India.
The author is thankful to Prof. Bei Lok Hu for helpful comments and to Kinjalk
Lochan for useful discussions. 

\textbf{Conflict of Interest Statement:} None of the authors have conflict 
of interest for the work done in this article. 


\begin{thebibliography}{00}
\bibitem{bei1} Bei Lok Hu and Enric Verdaguer. Liv.Rev.Rel. 11, 1 (2008). 
\bibitem{phillip} Nicholas G Phillips and B.L.Hu., Phys.Rev.D 63, 104001.(2001).
\bibitem{eftek} A.Eftekharzadeh  et.al, Phys Rev D, 85, 044037, (2012).
\bibitem{mink} Rosario Martin, Enric Verdaguer, 465 Issue( 1-4) Phys Lett B
  (1999).
\bibitem{desit} Enric Verdaguer, Phys. Conf. Ser 314, 012008  (2011). 
\bibitem{desit1} Enric Verdaguer, IJMPD 20,5, (2011).
\bibitem{martin} Rosario Martin and Enric Verdaguer., Phys.Rev.D, 60,084008,
(1999).
\bibitem{bei3} Bei Lok Hu and Enric Verdaguer.,  Semiclassical and Stochastic
Gravity, Cambridge University Press.,(2020)
\bibitem{chad}Chad R.Galley, Phys.Rev.Lett., 110,174301 (2013). 
\bibitem{bei2} B.L.Hu, E.Verdaguer., Class.Quant.Grav., 20,6 (2003).
\bibitem{madsen} M.S.Madsen, Class. Quant Grav, 5, 627 (1988).
\bibitem{forger} M.Forger, J.Laartz and U.Schaper.,Commun. Math.Phys, 159,
319-328 (1994). 
\bibitem{wald} Robert M.Wald, Quantum Field Theory in Curved Spacetime
and Black Hole Thermodynamics., The Univ. of Chicago Press (1994).
\bibitem{satin1} Seema Satin., Phys.Rev D, 51,4 (2019).
\bibitem{satin2} Seema Satin, Class. Quant. Grav, 39, 095004 (2022).
\bibitem{satin3} Seema Satin, IJMPD, 32,05 (2023).
\bibitem{sukanya} B.L.Hu and Sukanya Sinha, Phys.Rev.D, 51,4 (1994).
\bibitem{bei4} Esteban A.Calzetta and Bei Lok Hu., Non-equilibrium Quantum 
Field Theory., Cambridge University Press (2009). 
\end{thebibliography}
\end{document}